\newcommand{\teff}{t_{\rm eff}}

\newcommand{\chieq}{\chi_{\rm eq}}
\newcommand{\ising}{Fe$_{0.50}$Mn$_{0.50}$TiO$_3$ }
\newcommand{\AgMn}{{\textbf{Ag}}Mn }

\newcommand \be {\begin{equation}}
\newcommand \ee {\end{equation}}

\newcommand{\eq}[1]{Eq.~(\ref{#1})}
\newcommand{\kb}{k_{\rm B}}

\documentclass[aps,prl,twocolumn,superscriptaddress]{revtex4}

\usepackage{graphicx}

\begin{document}

\title{
Domain growth by isothermal aging in 3d Ising and Heisenberg spin glasses
}
\author{
P. E. J{\"o}nsson
}
\affiliation{
Department of Materials Science, Uppsala University, 
Box 534, SE-751 21 Uppsala, Sweden 
}
\author{
H. Yoshino
}
\affiliation{
Department of Earth and Space Science, Faculty of Science,
Osaka University, Toyonaka, 560-0043 Osaka, Japan
}
\author{
P. Nordblad
}
\affiliation{
Department of Materials Science, Uppsala University, 
Box 534, SE-751 21 Uppsala, Sweden 
}
\author{H. Aruga Katori}
\affiliation{
RIKEN, Hirosawa 2-1, Wako, Saitama, 351-0198, Japan
}
\author{A. Ito}
\affiliation{
RIKEN, Hirosawa 2-1, Wako, Saitama, 351-0198, Japan
}

\date{\today}

\begin{abstract}
Non-equilibrium dynamics of three dimensional model spin glasses -- 
the Ising system Fe$_{0.50}$Mn$_{0.50}$TiO$_3$ 
and the Heisenberg like system Ag(11 at\% Mn) -- has been investigated by 
measurements of the isothermal time decay of the low frequency ac-susceptibility 
after a quench from the paramagnetic to the spin glass phase. It is found that 
the relaxation data measured at different temperatures can be scaled  
according to predictions from the droplet scaling model,
provided that critical fluctuations are accounted for in the analyzes.

\end{abstract}

\maketitle

A spin glass ages --  the magnetic response of 
the system depends on its thermal history and especially the length 
of time it has been kept at a constant temperature in the glass phase \cite{lunetal}.
An aging phenomenon is also common to many other disordered and 
slowly relaxing physical systems \cite{struik-hametal-jonetal-albetal}. 
A thorough understanding of the driving forces behind
the aging process is thus important to adequately model glassy matter. 
Aging in spin glasses (SG) is within real space SG models 
associated with   
a slow growth of the coherence length for equilibrium SG order. The droplet 
theory \cite{fishus88noneq} provides concrete predictions of the scaling 
properties of the age dependent macroscopic dynamical
susceptibility in terms of a growing coherence length $L_{T}(t)$. 
One key test of the applicability of real space SG 
models is thus to quantitatively confront age dependent susceptibility data with 
the droplet model predictions. A positive result of such a study 
does not only give numbers for 
the parameters involved but allows further 
quantitative investigations of important ingredients of real space pictures such as
chaos with temperature and overlap length, and could even provide hints on the 
topological character of the assumed SG equilibrium domains. It does on the other hand 
not tell anything about the relevance of other approaches to spin glass physics.   
In this Letter, we investigate the time decay of the low frequency 
ac-susceptibility at constant temperature of two model spin glasses -- the 3d Ising 
system \ising and the 3d Heisenberg-like Ag(11 at\% Mn) (\textbf{Ag}Mn). 
In contrast to the outcome of all earlier corresponding studies 
(see for example \cite{sgexp_saclay,kometal2000}), we here find that the relaxation 
data can be scaled according to the explicit relation between
the growth of $L_{T}(t)$ and the relaxation of the ac-susceptibility 
suggested by the droplet theory \cite{fishus88noneq}, provided however
that critical fluctuations are accounted for in the analyzes. 

In the droplet model it is assumed that isothermal aging at a constant 
temperature $T$ after a quench  from a temperature above the spin-glass
transition temperature $T_{\rm g}$ is a coarsening 
process of domain walls as in many other phase ordering systems.
During the isothermal aging of a spin glass of spatial dimension $d$, the 
temporal ac-susceptibility at a given angular
frequency $\omega$ at time $t$ after the quench is supposed to scale as
\cite{fishus88noneq,Schetal93}, 
\begin{equation}
\frac{\chi''(\omega,t)-\chieq''(\omega)}{\chi''(\omega,t)}
\propto
\left[
\frac{L_T(1/\omega)}{L_T(t)}
\right]^{d-\theta} \, ,
\label{Eq: chi_noneq-droplet}
\end{equation}
where $L_T(t)$ is the typical size of the domains and $L_{T}(1/\omega)$ is
the typical size of the droplet being polarized by the ac field. The basic
physical idea behind this scaling form is that the presence of a domain wall
effectively reduces the excitation gap $\Upsilon(T)(L/L_{0})^{\theta}$
of droplet excitations that are touching the domain wall.
Here $\theta$ is  the stiffness exponent and $L_{0}$ is 
a microscopic length scale. 
The equilibrium ac susceptibility $\chieq''(\omega)$ 
is proposed to scale as \cite{fishus88eq},
\begin{equation}
\chieq''(\omega) \approx \frac{\pi}{2}
K_\omega \frac{q_m(T)}{\Upsilon(T)}
\frac{\partial}{\partial \ln \omega}
\left[\frac{L_{0}}{L_{T}(1/\omega)}
\right]^{\theta} \, ,
\label{Eq: chi_eq-droplet}
\end{equation}
where $K_\omega$ is an universal constant and $q_m$ is equal to the Edwards-Anderson order parameter for a symmetric spin glass.
The dynamics is governed
by thermally activated processes and the length scale $L$
associated with a certain temperature $T$ and the time scale $t$ is given by
\begin{equation}
L_T(t) 
\sim \left[ \frac{\kb T \ln (t/\tau_{0}(T))}{\Delta(T)} \right]^{1/\psi} \,,
\label{Eq: L-log}
\end{equation}
with $\tau_{0}(T)$ being the unit time scale for activated processes.
This growth law is derived from scaling of the typical free-energy 
barrier associated with
the length scale $L$, $F_{\rm barrier}(L) \sim \Delta(T) (L/L_{0})^{\psi}$
with the exponent $\psi >0$.
The stiffness for the excitation gap $\Upsilon(T)$ and barrier 
$\Delta(T)$ depends on temperature as 
$\Upsilon (T) \sim J\epsilon^{\theta\nu}$ and 
$\Delta (T) \sim J\epsilon^{\psi\nu}$, respectively, 
with $J$ being the characteristic energy scale and $\epsilon = T/T_g - 1$.

An important point 
is that the unit time scale $\tau_{0}(T)$ for activated hopping processes
is {\it not} simply given by the real microscopic time scale \cite{Risken}, 
which is $\tau_{\rm m} \sim \hbar/J \sim 10^{-13}$ s in spin systems. 
A plausible choice for $\tau_{0}(T)$ is instead the critical correlation time 
$\tau_c$ as proposed in \cite{fishus88noneq}, 
\be
\tau_{0}(T)\sim \tau_c 
\sim  \tau_{\rm m} (\xi(T)/L_{0})^{z} \sim
\tau_{\rm m} |\epsilon|^{-z\nu} 
\label{eq-tauT}
\ee 
with $z$ and $\nu$ being the dynamical critical exponent and the exponent
for the divergence of the correlation length $\xi(T)$, respectively.

Due to a possible slow crossover from critical dynamics
at $t \ll \tau_{c}(T)$ to activated dynamics at $t \gg \tau_{c}(T)$,
the logarithmic domain growth law \eq{Eq: L-log} may apply only for ideally 
asymptotic regimes 
beyond the time scales of numerical 
simulations \cite{Hukuetal2000}.
Results from numerical simulations \cite{MC-lt} accordingly suggest an alternative 
growth law for the spin glass coherence length
\begin{equation}
L_T(t) 
\sim
L_0(t/\tau_m)^{1/z(T)}
\label{Eq: L-alg}
\end{equation}
where $z(T)$ is a temperature-dependent effective exponent.
Inserting \eq{Eq: L-alg} into the basic scaling
forms \eq{Eq: chi_noneq-droplet} and \eq{Eq: chi_eq-droplet} one can
obtain corresponding formulas for the relaxation and frequency dependence of $\chi''$ 
\cite{kometal2000},
\begin{equation}
\frac{\chi''(\omega,t)-\chieq''(\omega)}{\chi''(\omega,t)}
\propto
(\omega t)^{-k(T)}\, ,   \qquad 
\chieq''(\omega,T) \propto \omega^{\alpha(T)}  
\label{Eq: X-wt}
\end{equation}
with temperature dependent effective exponents
\be
k(T) =  (d-\theta)/z(T)\, , \qquad \alpha (T)=  \theta / z(T) \, .
\label{eq-kt-alpha}
\ee
Interestingly, \eq{Eq: X-wt} is very close 
to that of the $\omega t$ -scaling proposed 
in previous experiments \cite{sgexp_saclay} where indeed the exponents
are found to depend on temperature. 

\begin{figure}[ht]
\includegraphics[width=0.5\textwidth]{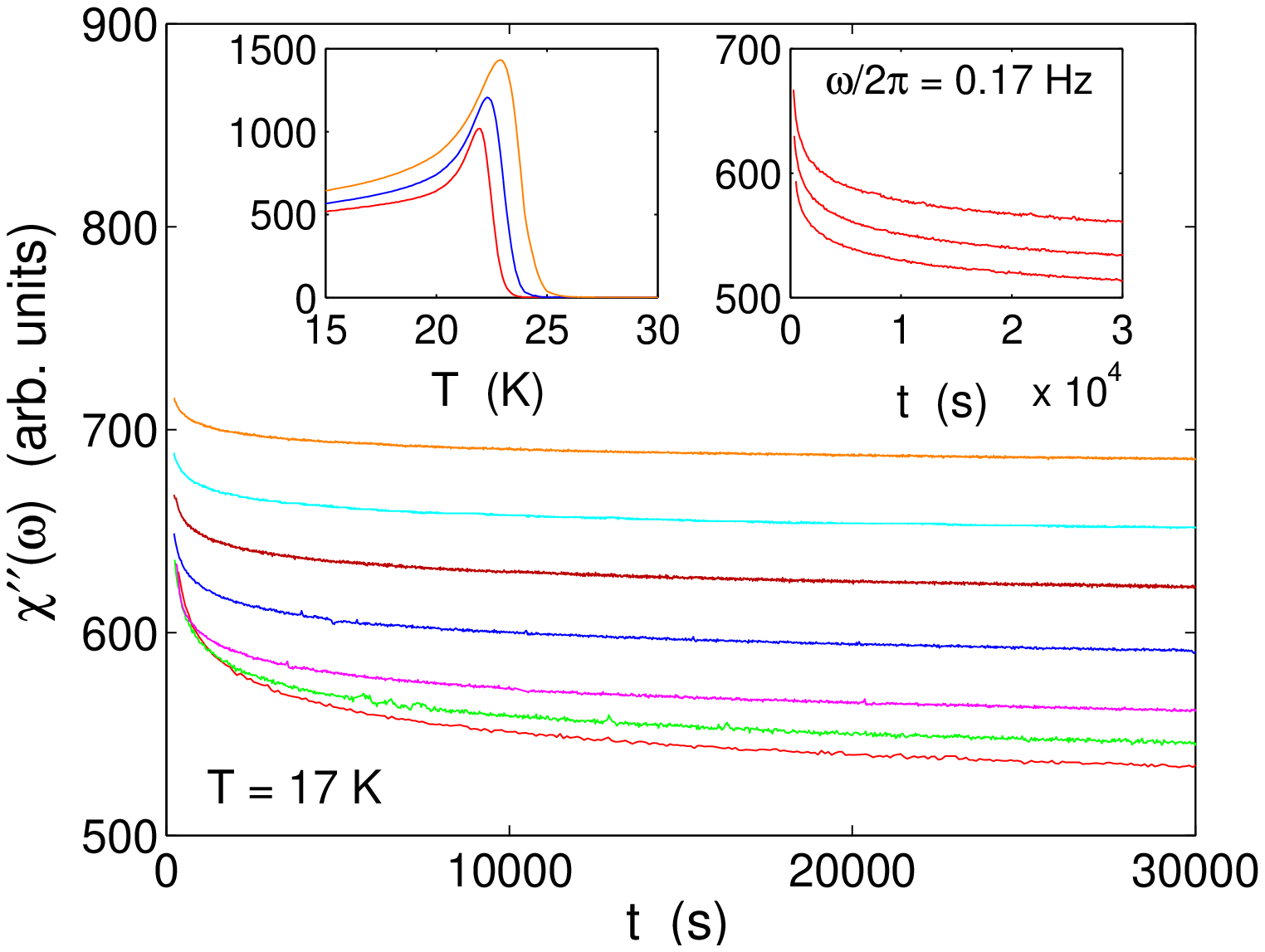}
\caption{
$\chi''(\omega) $ vs time for the \ising sample with $\omega / 2 \pi =0.17,0.51,1.7,5.1,17,55,170$ Hz (from bottom to top).
The left inset show $\chi''(\omega)$ vs temperature, measured on cooling, for $\omega / 2 \pi=0.17, 5.1$, and 170 Hz (from bottom to top). The right inset show $\chi''(\omega)$ vs time for $T=15, 17$, and 19 K (from bottom to top).
\label{Fig: xt}}
\end{figure}

The experiments were performed on a single crystal of the
short-range \ising Ising spin glass \cite{itoetal} 
and a polycrystalline sample of the long-range \AgMn Heisenberg-like spin glass.
The isothermal aging was measured by recording the ac-susceptibility, as a function of 
time (for up to $10^4$ s), after rapidly cooling the spin glass from a temperature above 
$T_g$ to the measurement temperature. 
The amplitude of the probing ac field 
was 0.02 Oe and frequencies in the range 
$\omega/2\pi=0.17$ -- 170 Hz were used. The background field was smaller than 0.1 mOe. 
The cooling 
rate, 0.05--0.08 K/s, was the fastest allowed by the non-commercial squid magnetometer 
\cite{magetal97lockin} employed in these studies. Some typical experimental results 
for the \ising sample are shown in Fig. \ref{Fig: xt}.
The relaxation toward equilibrium is for both samples so slow, 
even for the highest measurement frequencies, that 
$\chieq''(\omega)$ could not be estimated directly but had to be included as a 
fitting parameter in the analyzes at all frequencies and temperatures.
Due to the finite cooling rate 
the system will have an ``effective age'' when 
it has been cooled to the measurement temperature.
An ``effective time'' $\teff$ has therefore been added to the measurement 
time in the analyzes.
In the following analyzes 
we have used $\teff(T=30, 25,$ 20 K$)=10, 15,$ and 30 s for  the \AgMn 
sample and $\teff(T=19, 17,$ and 15 K$)=150, 200,$ and 350 s for the
\ising sample \cite{comment}.  

\begin{figure}[ht]
\includegraphics[width=0.5\textwidth]{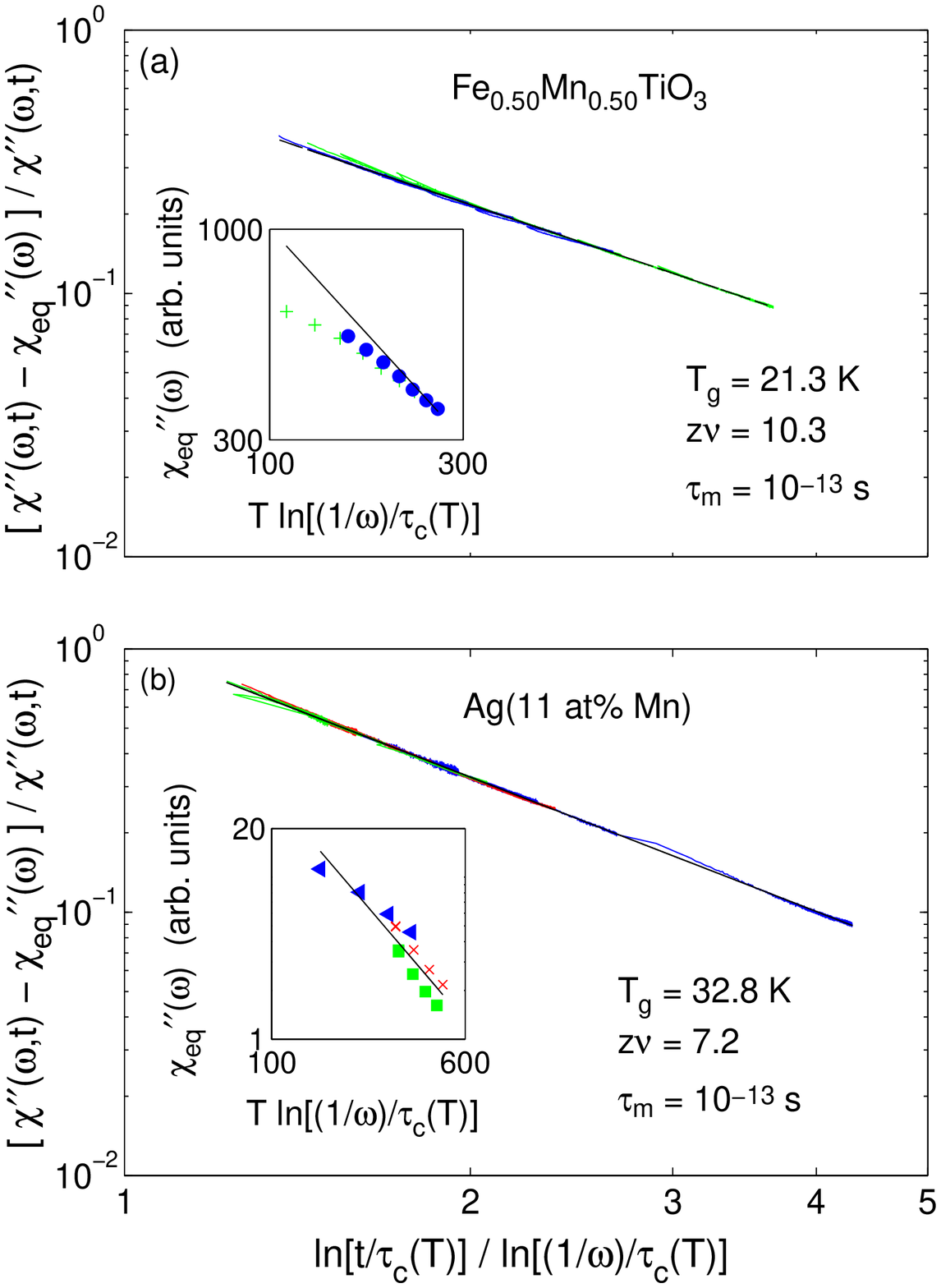}
\caption{
$[\chi''(\omega,t) - \chieq''(\omega)]/\chi''(\omega,t)$ vs $\ln[t/\tau_c(T)] / \ln[(1/\omega)/\tau_c(T)]$ on a log-log scale (a) for the \ising sample with $\omega / 2 \pi =0.17,0.51,1.7,5.1,17,55,170$ Hz, $T=15,17$ K and (b)  for the \AgMn sample with $\omega/2\pi = 0.17,1.7,17,170$ Hz, $T=20,25,30$ K.
The insets show $\chieq''(\omega)$ vs $\ln[(1/\omega)/\tau_c(T)]$ on a log-log scale (a) at $T= 17$ K (pluses), and 15 K (circles) and (b) $T= 30$ K (triangles), 25 K (crosses), and 20 K (squares).
\label{Fig: logt}}
\end{figure}

The isothermal aging data were first analyzed assuming the logarithmic domain growth law in \eq{Eq: L-log}, which inserted in \eq{Eq: chi_noneq-droplet} yields,
\begin{equation}
\frac{\chi''(\omega,t)-\chieq''(\omega)}{\chi''(\omega,t)}
\propto
\left\{
\frac{\ln[t/\tau_c(T)]}{\ln[(1/\omega)/\tau_c(T)]}
\right\}^{\frac{d-\theta}{\psi}} \, .
\end{equation}
This scaling law should permit all isothermal aging data from
measurements at different temperatures and frequencies to collapse onto
one {\it single} master curve.
A necessary condition for such a scaling is activated dynamics, i.e.  $t > (1/\omega) \gg \tau_c(T)$.
This condition is not fulfilled at 
$T=19$~K for the \ising sample, where $\tau_c=9\cdot10^{-4}$ s which is equal to $1/\omega$ for the highest frequency. The data at $T=19$ K have therefore been excluded
from this analysis.
The parameters describing the critical dynamics were varied 
within reasonable empirical values ($T_g = 32.5 - 33$~K, $z \nu = 7 - 8$, 
and $\tau_m \approx 10^{-13}$~s, for the \AgMn sample; $T_g=21.0 - 21.5$~K, 
$z\nu=10-11$, and $\tau_m \approx 10^{-13}$~s for the \ising sample). 
For the \AgMn sample, a good scaling could be obtained using critical    
parameters in the indicated range 
resulting in  $(d-\theta)/\psi \approx 1.7$ [c.f. Fig.~\ref{Fig: logt}(b)].
For the \ising sample the quality of the scaling was more sensitive to the 
parameters determining the critical dynamics.
However, a reasonably good scaling was obtained and 
we can estimate that $(d-\theta)/\psi \approx 1.5 $ 
[c.f. Fig.~\ref{Fig: logt}(a)]. 
It was not possible to scale the data 
for any of the two spin glasses using a constant value for  $\tau_0 \sim 10^{-13}$~s. 
The remaining fitting parameter in the scaling analysis, $\chieq''(\omega,T)$, 
is predicted to show a frequency and temperature 
dependence according
to \eq{Eq: chi_eq-droplet}, i.e. $\chieq''(\omega) \propto [T \ln((1/\omega)/\tau_c(T))]^{-(1+\theta/\psi)}$.
Asymptotic fits of the obtained $\chieq''(\omega,T)$ toward the
highest values of $T \ln[(1/\omega)/\tau_c(T)]$ to this equation then give 
$1+\theta/\psi \approx 1.8$ 
for the \AgMn sample and $1+\theta/\psi \approx 1.1$ for the \ising sample. 
Combining the results for the exponents we obtain $\psi \approx 1.2$ 
and $\theta \approx 1.0$ 
for the \AgMn sample while $\psi \approx 1.9$ and 
$\theta \approx 0.2$ for the \ising sample.
These estimates are within the imposed limits $\theta \le (d-1)/2$ and 
$\theta \le \psi \le d-1$ \cite{fishus88eq}. For both samples, the exponent $(d-\theta)/\psi$ is strongly dependent on the parameters describing the critical dynamics and 
increases with decreasing $\tau_c(T)$ (i.e. higher $T_g$ and/or smaller $z\nu$), 
while the exponent $1+\theta/\psi$ was less sensitive.
The here obtained value of $\psi$ for the \AgMn is close to a value reported earlier
$\psi \approx 1.3$ \cite{saclay}, and the extracted value of $\theta$ for the Ising spin glass is in
accord with estimations from numerical simulations on the 3d
Edwards-Anderson model \cite{theta,kometal2000}. 
On the other hand, for the Ising spin glass $\psi \approx 1.9$ is much 
larger than earlier reported experimental values 
$\psi \approx 0.8$ \cite{matetal95}
and $\psi \approx 0.3-0.7$ \cite{saclay}.   

To give some further perspective on the values obtained for the 
exponents $\theta$ and $\psi$, it should be pointed out that 
employing the alternative form  
$[\chi''(\omega,t) - \chieq''(\omega)]/\chieq''(\omega)$ 
on the right hand side of \eq{Eq: chi_noneq-droplet} yields an 
almost equally good scaling behavior, but that the values 
of the exponents alters to $\psi \approx 1.3$ and 
$\theta \approx 0.13$ for the \ising system and $\psi \approx 0.8$ and
$\theta \approx 0.6$ for the \AgMn sample.       
Here it is worth to note that the alternative
scaling form and the scaling form of \eq{Eq: chi_noneq-droplet} 
are expected to give the same result if correction terms at order 
$(L_{T}(1/\omega)/L_{T}(t))^{2(d-\theta)}$ are negligible.
However the ratio $L_{T}(1/\omega)/L_{T}(t)$
cannot reach values much smaller than $1$ 
on experimental time scales so that neglecting 
this term might have affected the analyzes. 
Also, the possible slow crossover from critical to activated dynamics may 
require a more complete functional form than only the asymptotic form
\eq{Eq: L-log} to precisely describe the growth law 
on experimental time scales. 

\begin{figure}[ht]
\includegraphics[width=0.5\textwidth]{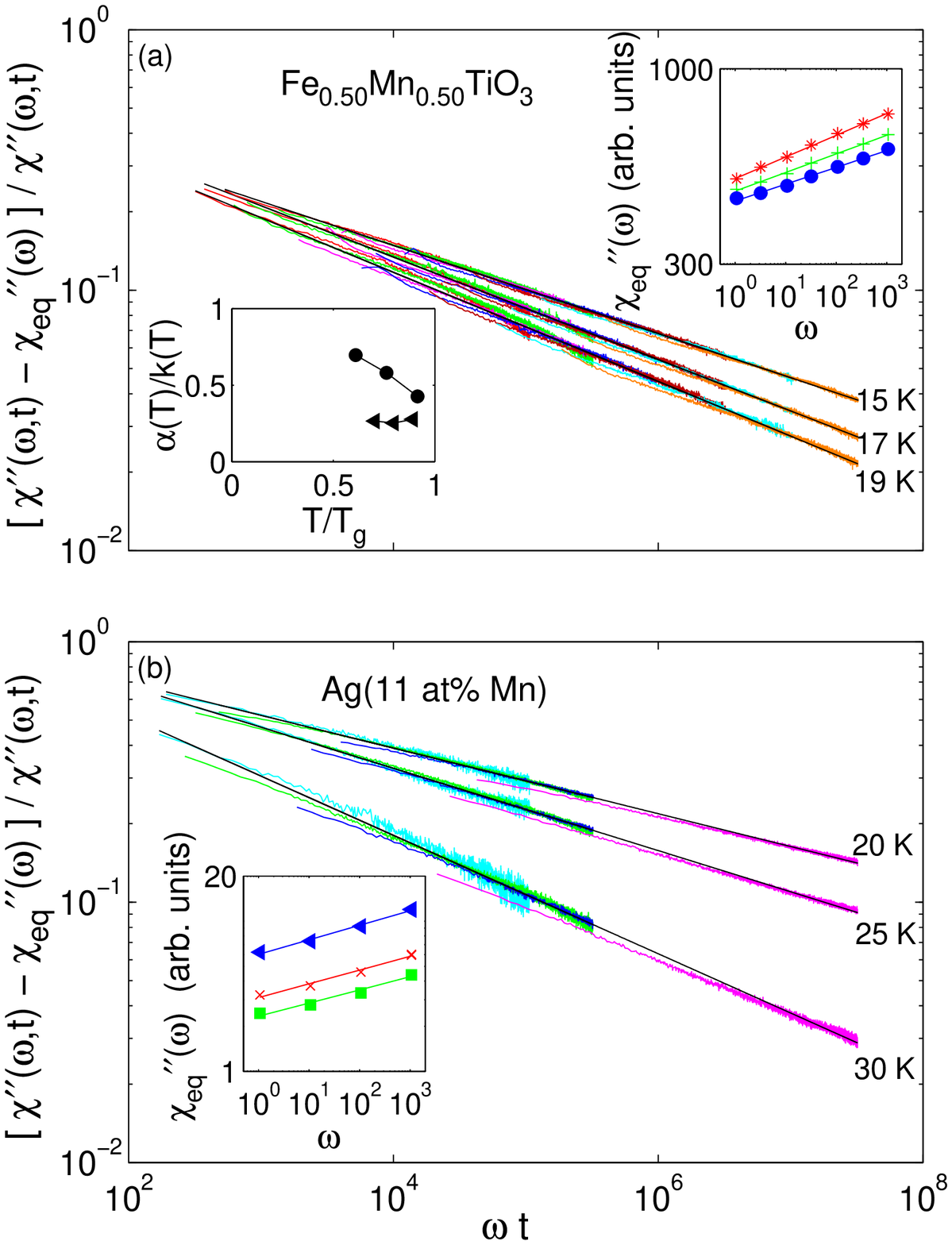}
\caption{
$[\chi''(\omega,t) - \chieq''(\omega)]/\chi''(\omega,t)$ vs $\omega t$ on a log-log scale (a) for the \ising sample with $\omega / 2 \pi =0.17,0.51,1.7,5.1,17,55,170$ Hz and (b)  for the \AgMn sample with $\omega/2\pi = 0.17,1.7,17,170$ Hz.
The left inset in (a) shows $\alpha(T)/k(T)$ vs $T/T_g$ for the \AgMn sample (circles) with $T_g=32.8$ K and for the \ising sample (triangles) with $T_g=21.3$ K.
The other insets show $\chieq''(\omega)$ vs $\omega$ on a log-log scale (a) at $T= 19$ K (stars), $T= 17$ K (pluses), and 15 K (circles) and (b) $T= 30$ K (triangles), 25 K (crosses), and 20 K (squares).
\label{Fig: wt}}
\end{figure}

Assuming instead that the domain growth is correctly described by the
algebraic growth law \eq{Eq: L-alg}, the isothermal aging should
scale according to the power law in \eq{Eq: X-wt} with exponents defined in \eq{eq-kt-alpha}. In this case, the data corresponding to different frequencies could 
be collapsed into one 
master curve for each temperature (see Fig.~\ref{Fig: wt}), using $\chi_{eq}''(\omega,T)$ as fitting parameter.
The $\chieq''(\omega)$ values obtained are shown in log-log plots in the insets 
of the figure together with fits to the power law behavior predicted in 
\eq{Eq: X-wt}. The 
temperature dependence of the exponent $k(T)$ [c.f. \eq{eq-kt-alpha}]
accords roughly with the behavior suggested by numerical simulations
$z(T) \simeq z (T_{\rm g}/T)$ \cite{MC-lt}.
However, some ambiguity appears when examining
the ratios $\alpha(T)/k(T)$ of the derived exponents 
(see inset of Fig.~\ref{Fig: wt}). At least for \AgMn this ratio clearly
increases with decreasing temperature,
while \eq{eq-kt-alpha} implies that it should be constant and equal to 
$\theta/(d-\theta)$. In the case of \ising the derived $\theta \approx$ 0.65, 
whereas for \AgMn $\theta$ appears to increase from about 
0.9 at 30 K to 1.2 at 20 K, which is outside the imposed limit $\theta<1$.

It has, as mentioned in the introduction, earlier been found that non-equilibrium 
relaxation in spin glasses cannot be described by $\ln(t)$-scaling
(see for example Refs. \cite{sgexp_saclay,kometal2000}).
However, in those investigations 
it was always assumed that $\tau_0(T) = \tau_m$. Quite remarkably, we now find that, 
accounting for critical fluctuations by identifying 
$\tau_0(T)$ with the critical correlation time, the $\ln(t)$-scaling 
permits to collapse all data measured at different frequencies and temperatures
onto one single master curve (see Fig. \ref{Fig: logt}). 
The importance of including critical fluctuations when analyzing non-equilibrium dynamics 
in spin glasses has also been emphasized 
in recent experimental work by E. Vincent and co-workers \cite{saclay}.
It can further be mentioned that numerical studies of the 4d Edwards-Anderson 
model only allow a logarithmic growth law if   
critical fluctuations are accounted for \cite{Hukuetal2000}.

The current finding, that 
the values of the exponents $\theta$ and $\psi$ for the 
\ising Ising spin glass are significantly different from those of 
the \AgMn Heisenberg-like spin glass, is consistent with 
the chiral glass picture proposed by Kawamura \cite{Kawamura}, conjecturing different 
universality classes for Heisenberg and Ising spin glasses, and numerical 
simulations within that picture indicate, in accord with our results, that 
the value of $\theta$ should be larger for a vector spin glass than for an Ising 
system \cite{maugre98}. Further experimental support for the existence of different 
universality classes in spin glasses was recently reported from magnetic torque 
measurements, yielding significantly different critical 
exponents for isotropic and anisotropic spin glasses \cite{petetal-pre}.

The numerical values of exponents involved in the growth law of 
the coherence length that have been obtained in this study 
will allow quantitative investigations of other aspects of spin glass 
dynamics within the droplet model, 
such as analyses of chaos with temperature and overlap length \cite{chaos}.

\begin{acknowledgments}
We thank Eric Vincent and Roland Mathieu for useful discussions.
This work was financially supported by the Swedish Research Council (VR).
\end{acknowledgments}


\begin{thebibliography}{10}
\bibitem{comment} 
The much longer values of $\teff$ required in the analyzes of 
the \ising sample than 
for the \AgMn sample is due to much stronger cooling rate effects in
the Ising than in the Heisenberg system.
\end{thebibliography}

\end{document}